\documentstyle[12pt]{article}
  \advance\oddsidemargin  by -1in
  \advance\evensidemargin by -1in
  \advance\topmargin      by -0.5in
\makeindex

\def\br{{\bf r}}
\def\brp{{\bf r^{\prime}}}
\def\Pom{{\bf I\!P}}
\def\lsim{\mathrel{\rlap{\lower4pt\hbox{\hskip1pt$\sim$}}
    \raise1pt\hbox{$<$}}}         
\def\gsim{\mathrel{\rlap{\lower4pt\hbox{\hskip1pt$\sim$}}
    \raise1pt\hbox{$>$}}}         
\def\beq{\begin{equation}}
\def\eeq{\end{equation}}
\def\bea{\begin{eqnarray}}
\def\eea{\end{eqnarray}}

\begin{document}

\begin{flushright}
{\em ITEP-PH-5/99\\
FZ-IKP(TH)-1999-32}
\end{flushright}
\vspace{1.0cm}

\begin{center}

{\Large \bf  The color dipole
 BFKL-Regge expansion:
 from DIS on protons
to pions to
rise of  hadronic cross sections 
\vspace{1.0cm}}\\

{\large \bf N.N.Nikolaev$^{1,2}$, J.Speth$^{1}$ and  V.
R.Zoller$^{2}$}\\

\vspace{0.5cm} $^{1}${ \em Institut  f\"ur
Kernphysik, Forschungszentrum J\"ulich,\\ D-52425 J\"ulich, Germany\\}

$^{2}${ \em L.D.Landau Institute for Theoretical Physics, Chernogolovka,
Moscow Region 142 432, Russia}\\

$^{3}${\em Institute for  Theoretical and Experimental Physics,\\
Moscow 117218, Russia\\
E-mail: zoller@heron.itep.ru}\\

\vspace{0.5cm}

{\bf Abstract}

\end{center}
As noticed by Fadin, Kuraev and Lipatov in 1975 incorporation of asymptotic
freedom
into the BFKL  equation splits the QCD pomeron into a  series of isolated poles
in
complex angular  momentum plane. Following our earlier work on the proton
structure
function  we explore the phenomenological consequences of the emerging
BFKL-Regge
 factorized
expansion for the small-$x$ structure function  of the pion $F_{2\pi}(x,Q^2)$.
 We calculate $F_{2\pi}$ in a
small-$x$ region and find good agreement with the recent  H1
determination of $F_{2\pi}(x,Q^2)$.
We also present the BFKL-Regge factorization based evaluation
 of the contribution from hard scattering to the observed rise
 of the $NN$,  $\pi N$ and real photo-absorption $\gamma N$ and $\gamma \pi$
 total cross sections.

\section{Introduction}

Recently there has been much discussion of the measurement of the pion
structure function $F_{2\pi}(x,Q^2)$ at HERA
 using deep inelastic scattering (DIS)
 of high-energy leptons on  the virtual pion of the
$\pi N$ Fock state of the proton and selecting semi-inclusive events
\beq
ep\to e^{\prime}nX
\label{eq:SEMINC}
\eeq
tagged  by a leading neutron  \cite{KOLYA,NSSS}, 
for earlier works see \cite{SULLIVAN}.
 The underlying DIS
\beq
\gamma(q) +\pi(k) \to X
\label{eq:GGTOX}
\eeq
is considered
in  the high-energy regime of very large Regge parameter
\beq
{1\over x}={W^2+Q^2\over Q^2}={1\over x_{Bj}}\gg 1,
\label{eq:LENGTH}
\eeq
where $W^2=(q+k)^2$ is the $\gamma^*\pi$ c.m.s.   collision
energy squared,
$Q^2=-q^2$ is the virtuality of the photon and
$x$ is the conventional Bjorken variable.
For $Q^2$ much larger
than the pion virtuality ($K^2=-k^2$) the pion can be considered
as the non-perturbative on-shell parton in the proton.
For the light-cone derivation of the flux of pions  in
 the $\pi N$  Fock state see
\cite{PIN,SPETHOM}, a similar flux is found in the recent Regge theory analysis
 \cite{NSSS}.
This  makes it possible to measure
$F_{2\pi}(x,Q^2)$   down to $x\sim 10^{-4}$ way beyond the reach of
$\pi N$ Drell-Yan experiment (hereafter in discussion of DIS we don't 
make distinction between $x$ and $x_{Bj}$).
Very recently the first experimental data on leading neutron production at HERA
and their interpretation in terms of $F_{2\pi}$ have been reported
\cite{F2PI}.

 Besides giving
access to the  $F_{2\pi}$, the semi-inclusive reaction (\ref{eq:SEMINC})
affects via unitarity the flavor content of the proton sea,
for the quantitative interpretation of the recent E866 data \cite{E866}
see \cite{NSSS}, for the review of earlier works see \cite{SPETHOM}.

In the literature there exist several parameterizations
 of parton distributions in the pion \cite{OLDPI,PIONGRV}
based on the $\pi N$ Drell-Yan data taken at relatively large
 $x\gsim 5\cdot 10^{-2}$.
Extrapolations of these parameterizations to $x\ll 1$ diverge quite strongly,
 which is not surprising if one recalls that the conventional DGLAP
phenomenology lacks
the predictive power at small $x$  and the small-$x$ extrapolations
 of the pre-HERA fits to the proton structure functions did generally miss
the small-$x$ HERA results \cite{HERA94}.

In this communication we address the issue of the small-$x$ behavior of
 the pion structure function in the BFKL-Regge framework.
 As noticed by Fadin, Kuraev and Lipatov \cite{FKL}
 and discussed in detail by Lipatov \cite{LIPAT86},
the incorporation of asymptotic freedom, i.e.  the
 running QCD coupling, into the BFKL equation makes
 the QCD pomeron a series of Regge poles.
 The contribution of the each pole to scattering
 amplitudes satisfies the standard Regge-factorization \cite{FACTORG}.
Recently we reformulated the BFKL-Regge expansion in the
 color dipole (CD) basis  \cite{JETPLett,DER}
which we apply here to calculation of the small-$x$
 behavior of  the pion structure function.
As a by-product of our analysis we present evaluations of the hard
scattering contribution to the observed rise of the pion-nucleon,
nucleon-nucleon  and real photo-absorption $\gamma N$
and $\gamma \pi$ total cross sections.

\section{The color dipole BFKL-Regge factorization}

In the color dipole basis the beam-target interaction amplitude
 is expanded in terms
of amplitudes of scattering of color dipoles $\bf r$ and
 $\bf{r^{\prime}}$ in both  the beam ($b$) and target ($t$) particles
 (here $\bf{r}$ and $\bf{r^{\prime}}$ are
the two-dimensional vectors in the impact parameter plane).
Invoking the optical theorem for forward scattering as
a fundamental quantity one can use
the dipole-dipole cross section
$\sigma(x,\bf{r},\bf{r^{\prime}})$.
The $\log{1\over x}$ BFKL evolution in the color dipole basis (CD BFKL)
 has been studied in detail in
\cite{JETPLett,DER,PISMA1,NZZJETP,NZHERA}.
Once $\sigma(x,\br,\brp)$
is known one can calculate the total cross section of $bt$ scattering
 $\sigma^{bt}(x)$
 making use of the
 color dipole factorization

\beq
\sigma^{bt}(x)
=\int dz d^{2}{\bf{r}} dz^{\prime} d^{2}{\bf{r^{\prime}}}
|\Psi_b(z,{\bf{r}})|^{2} |\Psi_{t}(z^{\prime},{\bf{r^{\prime}}})|^{2}
\sigma(x,{\bf{r},\bf{r^{\prime}}}) \,,
\label{eq:1.3}
\eeq
where $|\Psi_b(z,{\bf{r}})|^{2}$ and
$|\Psi_{t}(z^{\prime},{\bf{r^{\prime}}})|^{2}$
 are probabilities to find
 a color dipole,
$\br$ and $\brp$ in the beam and target, respectively.
Here we emphasize that the dipole-dipole cross section is beam-target symmetric
and
 universal for all beams and targets, all the beam and target dependence is
contained
in the color dipole distributions $|\Psi_b(z,\br)|^2$ and $|\Psi_t(z,\br)|^2$.

 We start
with the minor technical point that  $\sigma(x,\br,\brp)$ can depend
 on the orientation of the target and beam dipoles and can be expanded
 into the Fourier series
\beq
\sigma(x,\br,\brp)=\sum_{n=0}^{\infty}\sigma_n(x,r,r^{\prime})\exp(in\varphi)
\label{eq:ANGLE}
\eeq
where $\varphi$ is an azimuthal angle  between $\bf r$ and $\bf{r^{\prime}}$.
By the
nature of calculation of the beam-target total cross section $\sigma^{bt}(x)$
 only the term $n=0$ contributes in
(\ref{eq:1.3}).

 The CD BFKL-Regge factorization
uniquely prescribes that the $1/x$ dependence of the
 dipole-dipole total cross section
$\sigma(x,{r},{r^{\prime}})$
is of the form \cite{DER}
\beq
\sigma(x,r,r^{\prime})=\sum_{m}C_m\sigma_m(r)\sigma_m(r^{\prime})
\left({x_0\over x}\right)^{\Delta_m}\,.
\label{eq:PRODUCT}
\eeq
Here the dipole cross section $\sigma_m(r)$
 is an  eigen-function of the CD BFKL equation
\beq
{\partial\sigma_{m}(x,r)\over \partial \log(1/x)}=
 {\cal K}\otimes \sigma_{m}(x,r)=\Delta_{m}\sigma_{m}(x,r),
\label{eq:KO}
\eeq
 with
eigen value (intercept) $\Delta_{m}$ and $\sigma_{m}(x,r)$ being of the Regge
form
\beq
\sigma_{m}(x,r)=\sigma_{m}(r)\left(x_0\over x\right)^{\Delta_m}.
\label{eq:SIGRED}
\eeq
The running strong coupling exacerbates the well known infrared sensitivity
of the CD BFKL equation and infrared regularization is called upon:
infrared freezing of $\alpha_S$ \cite{GRIBOVLUND} and finite propagation radius $R_c$
of perturbative gluons were consistently used in our color dipole
approach to BFKL equation since 1994 \cite{PISMA1}. The past years the both concepts
 have become widely accepted: for a review of recent works on freezing
$\alpha_S$
see \cite{Dokshitzer}, our choice $R_c=0.27\,{\rm fm}$ has been confirmed by
the
recent determination of $R_c$ from the lattice QCD data on the field
 strength correlators \cite{MEGGI}.

\section{Summary on BFKL eigen-functions in the CD basis}

Here we recapitulate the principal findings on eigen-functions $\sigma_m(r)$
and
eigenvalues $\Delta_m$ of the CD BFKL equation
\cite{JETPLett,DER}. There is a useful similarity to solutions of the
Schr\"odinger
equation and the intercept plays the role of a binding energy.
  The leading eigen-function $\sigma_0(r)$ for the ground state with largest
binding energy $\Delta_0\equiv\Delta_{\Pom}$
 is node free.  The   eigenfunction $\sigma_m(r)$ for excited state  has
 $m$ radial  nodes.  With our infrared regulator
the intercept of the leading pole
  trajectory is  found to be  $\Delta_{\Pom}=0.4$\,.
 The intercepts $\Delta_m$  follow closely the law
$ \Delta_m= {\Delta_0/(m+1)}$ suggested first by Lipatov
from quasi-classical approximation to running BFKL equation in the related basis.
The sub-leading eigen-functions $\sigma_{m}(r)$  \cite{JETPLett} are
also similar  to Lipatov's quasi-classical eigen-functions \cite{LIPAT86} for
 $m\gg 1$.  For our specific choice of the infrared regulator
 the  node of $\sigma_{1}(r)$ is located at
$r=r_1\simeq 0.05-0.06\,{\rm fm}$, for larger $m$ the first nodes
move to a somewhat  larger $r$ and accumulate at $r\sim 0.1\, {\rm fm}$,
for the more detailed description of the nodal structure of $\sigma_m(r)$ see
\cite{JETPLett,DER}. Here we only emphasize that  for solutions with
$m\geq 4$
the  higher nodes are located at a very small $r$  way beyond the resolution
scale
$1/\sqrt{Q^2}$
 of foreseeable DIS experiments. Because for these higher solutions all
intercepts
$\Delta_m\ll 1$, in practical evaluation of $\sigma^{bt}$
 we can truncate expansion ($\ref{eq:PRODUCT}$) at $m=3$ lumping in the term
with $m=3$ contributions of all singularities with $m\geq 3$. Such a
truncation can be  justified {\sl a posteriori} if
such  a  contribution from $m\geq 3$ turns out to be a small
 correction, which will indeed be the case at very small $x$.

 The practical calculation of $\sigma(x,r,r^{\prime})$    by running the CD BFKL
evolution
requires the boundary condition $\sigma(x_0,r,r^{\prime})$ at certain $x_0\ll
1$.
The expansion coefficients $C_m$ in eq.(\ref{eq:PRODUCT}) are fully determined
by
 the boundary condition $\sigma(x_0,r,r^{\prime} )$ and by the choice
 of the normalization
of eigen-functions $\sigma_m(r)$. To this end
recall that the CD BFKL evolution sums the leading $\log(1/x)$
 diagrams for production of $s$-channel
gluons via exchange of $t$-channel perturbative gluons.
 It is tempting then, although not compulsory for any fundamental
 reasons, to take for boundary condition at $x=x_0$ the Born approximation,
 i.e. evaluate dipole-dipole scattering via the two-gluon exchange.
This leaves the starting point $x_0$ the sole parameter.
We follow the choice $x_0=0.03$
made in \cite{NZHERA}.
The very ambitious program of description of $F_{2}^{p}(x,Q^2)$
starting
 from  the, perhaps  excessively restrictive, but appealingly natural,
 two-gluon approximation has been launched by us in
\cite{JETPLett} and met  with  remarkable phenomenological success
\cite{DER}.

The exchange by perturbative gluons is a dominant mechanism for small dipoles
$r\lsim R_c$. In Ref.\cite{NZHERA,DER} interaction of large dipoles
has been modeled by the non-perturbative, soft mechanism which we
approximate here by a factorizable soft pomeron with intercept
$\alpha_{\rm soft}(0)-1=\Delta_{\rm soft}=0$, i.e., flat vs. $x$ at
small $x$. Then the extra term
$C_{\rm soft}\sigma_{\rm soft}(r)\sigma_{\rm soft}(r^{\prime})$
must be added in the  r.h.s. of expansion (\ref{eq:PRODUCT}). The
exchange by two non-perturbative gluons has been behind the
parameterization of $\sigma_{\rm soft}(r)$ suggested in \cite{NZHERA,JETPVM}
and used later on in \cite{JETPLett} and here, see also Appendix.
More recently several related models for $\sigma_{\rm soft}(r)$ have
appeared in the literature, see for instance models for dipole-dipole
scattering via polarization of non-perturbative QCD vacuum
\cite{Nachtmann}
and the model of soft-hard two-component pomeron \cite{LANDSH}.

\section{CD BFKL-Regge expansion for structure function}

Now we recall briefly the formalism for calculation of the target (t)
 structure
function ($t=p,\pi,\gamma^*,...$). It is convenient to introduce  the  eigen
structure functions $(m={\rm soft},0,1,2,...)$
\beq
f_m(Q^2)= {Q^2\over
 {4\pi^2\alpha_{em}}}
 \sigma_m^{\gamma^*}(Q^2)\, ,
\label{eq:F2CC}
\eeq
where
\beq
\sigma_m^{\gamma^*}(Q^2)=
 \langle {\gamma^*_T}  | \sigma_m(r)|{\gamma^*_T}
\rangle +  \langle {\gamma^*_L}  | \sigma_m(r)|{\gamma^*_L}
\rangle\,.
 \label{eq:SIGGPN}
\eeq
 Then the virtual $\gamma^{*}t$ photo-absorption cross section and the
target structure function $F_{2t}(x,Q^{2})$ take the form 
($m={\rm soft},0,1,2,...$)
\bea
\sigma^{\gamma^* t}(x,Q^2)=\sum_{m} C_m\sigma^{\gamma^*}_m(Q^2)\sigma^{t}_m
\left({x_0\over x}\right)^{\Delta_m}+\sigma^{\gamma^* t}_{\rm val}(x,Q^2)\\ \nonumber
=
\sum_{m} A^t_m\sigma^{\gamma^*}_m(Q^2)\left({x_0\over x}\right)^{\Delta_m}
+\sigma^{\gamma^* t}_{\rm val}(x,Q^2)
\,,
\label{eq:PRODUCT1}
\eea
\beq
F_{2t}(x,Q^2)=\sum A^{t}_m
f_m(Q^2)
\left({x_0\over {x}}\right)^{\Delta_m}
+F_{2t}^{\rm val}(x,Q^2)\,,
\label{eq:F2TRUN}
\eeq
where
\beq
\sigma^{t}_{m}=\langle t|\sigma_m(r)|t\rangle=
\int dz d^{2}{\bf{r}}
 |\Psi_{t}(z,r)|^{2}
\sigma_m(r) \,.
\label{eq:CN}
\eeq

The color dipole distributions in the transverse (T)
and  longitudinal (L)
photon of virtuality $Q^{2}$  derived in \cite{NZ91}
read

\beq
\vert\Psi_{T}(z,r)\vert^{2}={6\alpha_{em} \over (2\pi)^{2}}
\sum_{1}^{N_{f}}e_{f}^{2}
\{[z^{2}+(1-z)^{2}]\varepsilon^{2}K_{1}(\varepsilon r)^{2}+
m_{f}^{2}K_{0}(\varepsilon r)^{2}\}\,\,,
\label{eq:7.1.1}
\eeq
\beq
\vert\Psi_{L}(z,r)\vert^{2}={6\alpha_{em} \over (2\pi)^{2}}
\sum_{1}^{N_{f}}4e_{f}^{2}\,\,
Q^{2}\,z^{2}(1-z)^{2}K_{0}(\varepsilon r)^{2}\,\,,
\label{eq:7.1.2}
\eeq
where
\beq
\varepsilon^{2}=z(1-z)Q^{2}+m_{f}^{2}\,\,\,.
\label{eq:7.1.3}
\eeq
In Eqs.~(\ref{eq:7.1.1})-(\ref{eq:7.1.3})  $K_{0}$ and $K_{1}$ - are the
modified Bessel functions, $e_f$ is the quark charge,
 $m_{f}$ is the quark mass, $\alpha_{em}$ is the fine structure constant
 and $z$ is the
Sudakov variable, i.e. the fraction of photon's light-cone momentum
carried by one of the quarks of the pair ($0 <z<1$).
The functional form of $f_m (Q^2)$ convenient in applications
was presented in \cite{DER} (see also Appendix).
For the practical phenomenology at moderately small $x$ we
include a contribution from DIS on valence  quarks in the target
$F_{2\pi}^{\rm val}(x,Q^2)$ which is customarily associated with
the non-vacuum reggeon exchange.

In the evaluation of the proton structure function $F_{2p}(x,Q^2)$
we use the symmetric oscillator wave function of the $3$-quark proton.
We recall that in this approximation the  proton looks as $3/2$
color dipoles spanned between quark pairs. The  distribution
of sizes of  dipoles ${\bf r}$ spanned between quark pairs in the proton reads
\beq
|\Psi_{p}(r)|^2={1\over {2\pi\langle r^2_{p}\rangle}}
\exp\left(-{r^2\over{2\langle r^2_{p}\rangle}}\right)
\label{eq:WFPROTON}
\eeq
where $\langle r^2_{p}\rangle = 0.658\, {\rm fm^2}$ as
suggested by the standard dipole form factor of the proton.
In \cite{DER} we introduced normalization of  $\sigma_m(r)$ such that
for the proton target $(m={\rm soft},0,1,2,...)$
\beq
A_m^p=C_m\sigma_m^{p}=1.
\label{eq:NORM}
\eeq
Within this convention we have the proton
expectation values $\sigma^p_m$ and  parameters
$C_m$ of the truncated,  $m={\rm soft},0,1,2,3$, dipole-dipole expansion
cited in the Table 1.

We recall that because of the diffusion in color dipole space,
 exchange by perturbative gluons contributes also to interaction of large dipoles
$r>R_c$ \cite{NZZJETP}. However at moderately large Regge parameter this hard 
interaction driven effect is still small. For this reason in what follows we
 refer to terms $m=0,1,2,3$ as hard contribution
as opposed to the genuine soft interaction. 

\vspace{0.25cm}
\begin{center}
Table 1. CD BFKL-Regge expansion parameters.\vspace{0.45cm}\\

\begin{tabular}{|c|c|c|c|c|c|c|c|} \hline
$m$& $ \Delta_m$ & $\sigma_m^{p},{\rm\, mb}$ & $\sigma^{\pi }_m,{\rm\, mb}$&
 $C_m  ,{\rm\, mb^{-1}}$& $A_m^{\pi}$& $A_m^p$&
 $\sigma_m^{\gamma^*}(0),{\rm\, \mu b}$\\ \cline{1-8}
0 & 0.40  & 1.243 & 0.822 &0.804 & 0.661  &1. &6.767   \\ \cline{1-8}
1 &0.220  &0.462 & 0.303 &2.166 & 0.656 &1.& 1.885 \\ \cline{1-8}
2 &0.148  &0.374 & 0.244 &2.674 & 0.653 &1.& 1.320   \\ \cline{1-8}
3 &0.111  &0.993 & 0.647 &1.007 & 0.651 &1.& 3.186  \\ \cline{1-8}
{\rm soft} & 0.&31.19      & 18.91      &0.0321  & 0.606  &1. &79.81 \\
\cline{1-8}
\end{tabular}
\end{center}

\section{CD BFKL-Regge predictions for $F_{2\pi}(x,Q^2)$}

The extension to small-$x$ DIS off pions   is quite straightforward.
In the normalization (\ref{eq:NORM})
\beq
A^{\pi}_m={\sigma_m^{\pi }\over
\sigma_m^{p}}\,.
\label{eq:RATIO}
\eeq
We evaluate $\sigma_m^{\pi}=  \langle \pi| \sigma_m(r)|\pi
\rangle$ with the oscillator approximation for
 the $q\bar q$ wave function of the pion
\beq
|\Psi_{\pi}(r)|^2={3\over {8\pi\langle r^2_{\pi}\rangle}}
\exp\left(-{3r^2\over{8\langle r^2_{\pi}\rangle}}\right)\, ,
\label{eq:WFPION}
\eeq
where the charge radius of the pion suggests
$\langle r^2_{\pi}\rangle = 0.433\, {\rm fm^2}$.
Then the calculation of  $\sigma_m^{\pi}$ is parameter-free,
the results for  $\sigma_m^{\pi}$ and $A^{\pi}_m$ are cited in Table 1.

The minor change is that in a scattering of color dipole on the
pion the effective dipole-dipole collision energy is $3/2$ of that
in the scattering of color dipole on the three-quark nucleon
which is our reference process at the same total c.m.s. energy $W$.
Then the further evaluation of $\sigma^{\gamma^*\pi}(x,Q^2)$ and of
the pion structure function
\beq
F_{2\pi}(x,Q^2)=\sum_{m=0}^3A^{\pi}_m
f_m(Q^2)
\left({3\over 2}{x_0\over {x}}\right)^{\Delta_m}
+
A^{\pi}_{\rm soft}f_{\rm soft}(Q^2)+F_{2\pi}^{\rm val}(x,Q^2)\,
\label{eq:F2TRPI}
\eeq
shown by the solid curve in fig.~1 does not involve any adjustable parameters.
The valence component of the pion structure function (the dot-dashed curve)
is quite substantial at $x\gsim 10^{-2}$, but it is reasonably
constrained by the $\pi N$ Drell-Yan data and can be regarded as weakly
model dependent. In our calculations we employ the recent parameterization
\cite{PIONGRV}. We find a good agreement with the recent H1  determination
of $F_{2\pi}(x,Q^2)$ \cite{F2PI}. We do not see any need to modify the 
above specified natural, perturbative two-gluon exchange, boundary condition 
to CD BFKL evolution.

Although the agreement with the experimental data is good, testing the 
complete CD BFKL-Regge expansion in all its complexity is not a trivial task.
As we mentioned in the Introduction, small-$x$ extrapolations of existing 
DGLAP fits \cite{OLDPI,PIONGRV} to the  $\pi N$ Drell-Yan data diverge
quite strongly, but the proper fine tuning of the input can produce the 
DGLAP fits which in a limited range of moderately small $x$ will be very 
similar to our results. To this end, several aspects of the predictive 
power of CD BFKL approach are noteworthy. First, there is an obvious 
prediction that the leading hard pole contribution to single-pomeron
exchange will dominate at sufficiently small $x$ for all $Q^{2}$, but it 
is not easy to check it within the reach of the HERA experiments and better
understanding of the unitarity corrections to single-pomeron exchange is
needed at extremely small $x$ beyond the HERA range. 
Second, as noticed in \cite{NZZJETP} because 
of the short propagation radius for perturbative gluons, $R_{c}^{2} \ll
\langle r_{p}^{2}\rangle,\langle r_{\pi}^{2}\rangle$, in our
CD BFKL approach  we have an approximate 
additive quark counting for hard components,
$m=0,1,2,3$, i.e.,
\beq
{\sigma_m^{p}\over \sigma^{\pi }_m} \approx {3\over 2}\,,
\label{eq:AQM}
\eeq
see entries in Table 1. Third, according to the conventional estimates
for the proton and pion radii based on the experimental data on charge
radii, the dipoles spanned between quark pairs in the proton, and the 
quark and antiquark in the pion, have approximately similar sizes, cf. 
(\ref{eq:WFPROTON}) and (\ref{eq:WFPION}). Consequently, for a reason 
quite different from that for the hard BFKL exchange, an
approximate additive quark counting holds also for the soft-pomeron exchange,
see Table 1. A good approximation to the CD BFKL-Regge factorization 
prediction is
\beq
F_{2\pi}(x,Q^2)\simeq {2\over 3}F_{2p}\left({2\over 3}x,Q^2\right)\, .
\label{eq:F2PIFAC}
\eeq
This is an arguably natural, but far from trivial, relationship. Given 
that the origin of additive quark counting for hard and soft components 
is so different, it is not surprising that (\ref{eq:F2PIFAC})  
has not necessarily been
borne out by the diversity of the pre- and post-HERA DGLAP fits to the 
pion and proton structure functions. Fourth, the finding of a substantial 
soft contribution shown by the short-dashed line must not surprise anyone 
in view of a familiar strong sensitivity of the results of DGLAP evolution 
to the input structure function at a semi-hard starting point. Fifth, 
the CD BFKL approach predicts uniquely, that sub-leading eigen structure 
functions  $f_{m\geq 1}(Q^2)$ have their first node at $Q^{2} \sim 20-60$ 
GeV$^{2}$, see \cite{JETPLett,DER} and Appendix, and we would like to
comment on this node effect in more detail. 

The point is that in the vicinity of the node of sub-leading contributions  
the pion structure function is well reproduced by the Leading Hard$+$ 
Soft$+$Valence Approximation (LHSVA)
\beq
F^{\rm LHSVA}_{2\pi}(x,Q^2)\simeq {2\over 3}\left[f_0(Q^2)
\left({3\over 2}{x_0\over {x}}\right)^{\Delta_0}
+
f_{\rm soft}(Q^2)\right]+F_{2\pi}^{\rm val}(x,Q^2) \,,
\label{eq:LHPSA}
\eeq
which gives a unique handle on the intercept $\alpha_{\Pom}(0)=1+\Delta_0$
of the leading hard BFKL pole. This point is illustrated in fig.~1
where the dotted curve represents the LHSVA. 
A comparison with the solid curve for the 
complete BFKL-Regge expansion shows that, as a matter of fact, the 
contribution from sub-leading BFKL-Regge poles is marginal at all 
values of $Q^{2}$ of the practical interest. In order to delineate 
the impact of the valence component we show by the long-dashed curve 
the sum of the soft and leading hard pole contributions without 
the valence component. Even such a crude two-pole model does a good 
job at $x \lsim $(2-5)$\cdot 10^{-3}$, although at larger $x$ the
impact of the valence component on the $x$-dependence of the pion
structure function is substantial. The accuracy 
of the LHSVA improves rapidly at larger $Q^{2}$, see the boxes for 
$Q^2=7.5$ to 13.3 to 28.6 and 100\,GeV$^2$ in fig.~1. As a  
manifestation of the nodal 
structure of $f_{m\geq 1}(Q^2)$ the LHSVA overshoots the result of 
the full complete-Regge expansion slightly at $Q^2=28.6$ and $100\,{\rm GeV^2}$, 
while it undershoots the complete BFKL-Regge expansion slightly at 
$Q^2 \leq 13.3\,{\rm GeV^2}$. 

As eq.~(\ref{eq:F2PIFAC} suggests, this discussion of LHSVA is fully 
applicable to the proton structure function as well. This point is
demonstrated by the decomposition
into different components of our earlier CD BFKL-Regge predictions  
for the proton structure function \cite{JETPLett,DER}
shown in Fig.~2. Because the LHSVA formula (\ref{eq:LHPSA}) is 
sufficiently simple and the soft contribution is small for $Q^{2} 
\sim $10-100 GeV$^{2}$, a reliable determination of $\alpha_{\Pom}(0)$ 
for the rightmost BFKL singularity is feasible even from the pion 
data. An independent determination of $\alpha_{\Pom}(0)$ and, consequently,
the test of the CD BFKL approach is possible from the charm structure 
function of the proton - as we argued elsewhere \cite{NZcharm}, it 
receives a negligible soft and sub-leading contributions and is entirely 
dominated by the contribution from the rightmost BFKL singularity
(for the related recent discussion of the charm structure function
see also \cite{DLcharm}).

\section{Extending CD BFKL-Regge expansion to real photons and hadrons}

Finally, with certain reservations on absorption corrections
we can extend the BFKL-Regge factorization
from DIS to real photo-production and even hadron-hadron scattering.
Take for instance pion-nucleon scattering. There is always a contribution
from small-size dipoles in the pion to the color dipole factorization
formula (\ref{eq:PRODUCT}). For example, a probability $w_{\pi}(r<r_0)$
to find dipoles of
size $r\lsim r_0$ can be estimated as
\beq
w_{\pi}(r<r_0)\sim {3\over 8}{r_0^2\over \langle r^2_{\pi}\rangle},
\label{eq:WPROB}
\eeq
which gives quite a substantial fraction of the pion,
\beq
w_{\pi}(r<0.2\,{\rm fm})\sim 3\cdot 10^{-2}
\label{eq:WNUM}
\eeq
the interaction of which with the target nucleon proceeds in the
legitimate hard regime typical of DIS. The corresponding contribution
to $\sigma_{\rm tot}^{\pi N}$ must exhibit the same rapid rise with energy as
the proton structure function. Furthermore, in the BFKL approach
there is always a diffusion in the dipole size by which there is
a feedback from hard region to interaction of large dipoles and
vice versa. How substantial is such a hard contribution to the
hadronic and real photon total cross sections?

The explicit realization of this idea within the CD BFKL-Regge factorization
is embodied in expansion for the vacuum exchange
 contributions to the total cross sections 
\beq
\sigma^{pp}=\sum_{m}A^{p}_m
\sigma^{p}_m
\left({2\over 3}{x_0\over x}\right)^{\Delta_m}\,,~~~~
\sigma^{\gamma p}=\sum_{m}A_m^p \sigma_m^{\gamma^*}(0)
\left({x_0\over x}\right)^{\Delta_m}\,,
\label{eq:PP}
\eeq

\beq
\sigma^{\pi p}=\sum_{m}A^{\pi}_m
\sigma^{p}_m
\left({x_0\over x}\right)^{\Delta_m}\,,
~~~~
\sigma^{\gamma \pi}=\sum_{m} A_m^{\pi}\sigma_m^{\gamma^*}(0)
\left({3\over 2}{x_0\over x}\right)^{\Delta_m}\,,
\label{eq:GAMPI}
\eeq
where $m={\rm soft},0,1,2,3$. In the real photo-production limit 
the light quark masses $m_{u,d}=0.135\, {\rm GeV},\, m_{s}=0.285\, {\rm GeV}$
 serve to define
the transverse size of the hadronic component of the photon which
it is reasonable to expect close to the
transverse size of vector mesons and/or pions.
The real photo-absorption cross sections $\sigma_m^{\gamma^*}(0)$ and
$\sigma^{\gamma^*}_{\rm soft}(0)$ as given by
eqs.(\ref{eq:F2CC},\ref{eq:FSOFT})
extended to $Q^{2}=0$ are cited in Table 1.
Regarding the similarity of the transverse size of real
 photon and $\rho/\pi$-meson, notice that the ratio 
$\sigma^{\gamma^*p}_{\rm soft}/\sigma^{\pi p}_{\rm soft}\sim 1/230$ is very close to
the standard vector meson dominance estimate \cite{BAUER}.

In eqs.(\ref{eq:PP},\ref{eq:GAMPI})
the plausible choice of the Regge parameter is
${1/ x}={W^2/ m^2_{\rho}}\,,$
 in the NN and $\gamma \pi$ scattering we introduce the aforementioned energy rescaling
factor $2/3$ and $3/2$, respectively.
In our approach $\Delta_{\rm soft}=0$ and
 $\sigma_{\rm soft}={\rm const}(W^2)$,
so that the intrusion of hard regime into soft scattering
 is the sole source
of the rise of total cross sections. The CD BFKL-Regge factorization based
evaluation of the vacuum exchange contribution
to  $pp,\pi p, \gamma p$ and  $ \gamma \pi$ total cross sections \cite{PDG}
is
presented in Fig.~3. At low energies the vacuum exchange
 contribution underestimates the observed total cross section which can
 evidently be attributed to the non-vacuum Regge exchanges not
 considered here.

Although the hard pomeron component is  important
in both the 
purely  hadronic ($NN,\pi N$) and
 photo-absorption reactions, there is an important distinction between
 the two cases. Namely, in contrast to the proton/pion wave function 
(\ref{eq:WFPROTON},\ref{eq:WFPION}) 
which is smooth at $r\to 0$, the real  photon wave function squared $|\Psi_{T}(r)|^2$  
is singular (\ref{eq:7.1.1}).
 Furthermore, this singularity is a legitimate pQCD effect and
makes (i) evaluation of hard contributions, $m=0,1,2,3$, to $\gamma\pi,\gamma p$ 
 cross sections more reliable and (ii) uniquely predicts that hard contribution to
 $\sigma_{\rm tot}^{\gamma\pi},\sigma_{\rm tot}^{\gamma p}$ is relatively stronger than to 
$\sigma_{\rm tot}^{\pi p},\sigma_{\rm tot}^{p p}$. Indeed,
a closer inspection of the Table 1 shows that
in $\gamma N$  the relative weight
  of  hard component $\sum_0^3\sigma_m/\sigma_{\rm soft}$
 is almost twice as large
as in the  $\pi N$ scattering. 

We observe that hard effects in both the
 $p p, \pi p$ and $\gamma\pi, \gamma p$  
 to exhaust to a large extent, or even completely,
the observed rise of $\sigma_{tot}(W^2)$ at moderately large $W^{2}$.
Furthermore, there arises an interesting scenario,  discussed first in
 a simplified version in \cite{KOPNIKPOT}, in which soft interactions 
at moderately high energies can be described by the soft pomeron with  
$\Delta_{\rm soft}=0$ plus small contribution from hard scattering, 
the combined effect of which mimics the usually discussed phenomenological
 pomeron pole endowed with the intercept $\Delta\simeq 0.08$ \cite{LANDSHOFF}.
In this scenario the real issue is not an explanation of the rise of hadronic
$\sigma_{tot}(W^2)$
at moderately high energies, rather it  is whether there exists a mechanism 
to tame a too rapid rise of extrapolation of the hard component of the
 CD BFKL-Regge expansion
to very high-energies. To this end recall that 
 we considered only the non-unitarized running CD BFKL amplitudes 
too rapid a rise of which
must be tamed by the unitarity absorption corrections.
 The discussion of the issue of unitarization
goes
beyond the scope of the present communication.

\section{Conclusions}

We explored the consequences for small-$x$ structure functions
 and high energy total cross sections from the color dipole BFKL-Regge 
factorization. We
use very restrictive  perturbative two-gluon exchange as a 
parameter-free boundary condition for  BFKL evolution in the 
color dipole basis. Under plausible assertions on the color dipole 
structure of the pion, our parameter-free description of
 the pion structure function  agrees well with the H1 determinations. 
The found 
relationship between $F_{2p}(x,Q^2)$ and $F_{2\pi}(x,Q^2)$ is very close to 
$
F_{2\pi}(x,Q^2)\simeq {2\over 3}F_{2p}({2\over 3}x,Q^2)$. 
The contribution of sub-leading BFKL-Regge poles to the pion and proton 
structure function is found to be small in a broad range of $Q^{2}$ 
of the practical interest, which makes feasible the determination of
the intercept $\alpha_{\Pom}(0)$ for the rightmost BFKL singularity 
form the structure function data.

Because of the diffusion in color dipole space, the hard scattering 
which is a dominant feature of $F_{2\pi}(x,Q^2)$ and $F_{2p}(x,Q^2)$
at large $Q^2$ contributes also to real photo-absorption and hadron-hadron scattering.
This make plausible a scenario in which the rising hard component and the genuine
soft component with $\Delta_{\rm soft}=0$ mimics the effective 
 vacuum pole with  $\Delta_{\rm soft}\simeq 0.08$.\\

{\bf Acknowledgments: } This work was partly supported by the grants
INTAS-96-597 and INTAS-97-30494 and DFG 436RUS17/11/99.
\newpage

{\large\bf Appendix}\\

Although we have certain ideas on the shape of eigen-functions 
$\sigma_m(r)$ as a function of $r$ and/or eigen structure functions 
$f_m(Q^2)$ as a function of $Q^2$ \cite{JETPLett}, they are only  available
as a numerical solution to the running color dipole BFKL equation.
 On the other hand, for the practical  
 applications it is convenient to 
represent the results of numerical solutions for $f_m(Q^2)$
 in an analytical form
\beq
f_0(Q^2)=
a_0{R_0^2Q^2\over{1+ R_0^2Q^2 }}
\left[1+c_0\log(1+r_0^2Q^2)\right]^{\gamma_0}\,,
\label{eq:F20}
\eeq

\beq
f_m(Q^2)=a_m f_0(Q^2){1+R_0^2Q^2\over{1+ R_m^2Q^2 }}
\prod ^{n_{max}}_{i=1}\left(1-{z\over z^{(i)}_m}\right)\,,\,\, m\geq 1\,,
\label{eq:FN}
\eeq
where $\gamma_0={4\over {3\Delta_0}}= {10\over 3}$ 
and
\beq
z=\left[1+c_m\log(1+r_m^2Q^2)\right]^{\gamma_m}-1 ,\,\,\,
\gamma_m=\gamma_0 \delta_m
\label{eq:ZFN}
\eeq
and $m_{max}=$min$\{m,2\}$.

The nodes of $f_m(Q^2)$ are spaced by 2-3 orders of magnitude in $Q^2$-scale.
The first nodes of sub-leading $f_{m}(Q^2)$
are located at $Q^2\sim 20-60\, GeV^2$, the second nodes of $f_{2}(Q^2)$ and
$f_{3}(Q^2)$ are at $Q^2\simeq 5\cdot 10^3\, GeV^2$ and
$Q^2\simeq 2\cdot 10^4\, GeV^2$, respectively.
The third node of $f_3(Q^2)$ is at $\sim 2\cdot 10^7\,
GeV^2$, way beyond the reach of accelerator experiments at small $x$.
 The parameters  tuned to reproduce
 the numerical results for $f_m(Q^2)$ at $Q^2\lsim 10^5\, GeV^2$
are listed in the Table 2.
 For $m=3$ in this limited range of $Q^2$
 we take a simplified form with only two
first nodes, it must not be used for $Q^{2} \gsim 10^5$ GeV$^2$.\\

\vspace{0.5cm}
\begin{center}
Table 2. CD BFKL-Regge structure functions parameters.\vspace{0.45cm}\\

\begin{tabular}{|l|l|l|l|l|l|l|l|} \hline
$n$ & $a_m$  & $c_m$ & $r_m^2\,,$ ${\rm GeV^{-2}}$ &
$ R_m^2\,,$ ${\rm GeV^{-2}}$ &
$z^{(1)}_m$ & $z^{(2)}_m$ & $\delta_m$ \\ \cline{1-8}
0 & 0.0232  & 0.3261&1.1204&2.6018& & & 1. \\ \cline{1-8}
1 & 0.2788 &0.1113&0.8755&3.4648&2.4773 &    &1.0915 \\ \cline{1-8}
2 & 0.1953 &0.0833&1.5682&3.4824 &1.7706 &12.991 &1.2450  \\ \cline{1-8}
3 &0.4713&0.0653&3.9567&2.7756    &1.4963 &6.9160  &1.2284 \\ \cline{1-8}
{\rm soft}&0.1077 & 0.0673& 7.0332 & 6.6447 &   &  &      \\  \cline{1-8}
\end{tabular}
\end{center}
\vspace{0.5cm}

The soft component of the proton structure function derived from eq.(\ref{eq:F2CC})
with $\sigma_{\rm soft}(r)$ taken from \cite{JETPVM} is parameterized as follows
\beq
f_ {\rm soft}(Q^2)= 
{a_{\rm soft}R^2_{\rm soft}Q^2\over{1+R^2_{\rm soft} Q^2 }}
\left[1+c_{\rm soft}\log(1+r^2_{\rm soft}Q^2)\right]\,,
\label{eq:FSOFT}
\eeq
with parameters cited in the Table 2.


\newpage

{\large\bf{ Figure Captions}\\ }

\begin{enumerate}

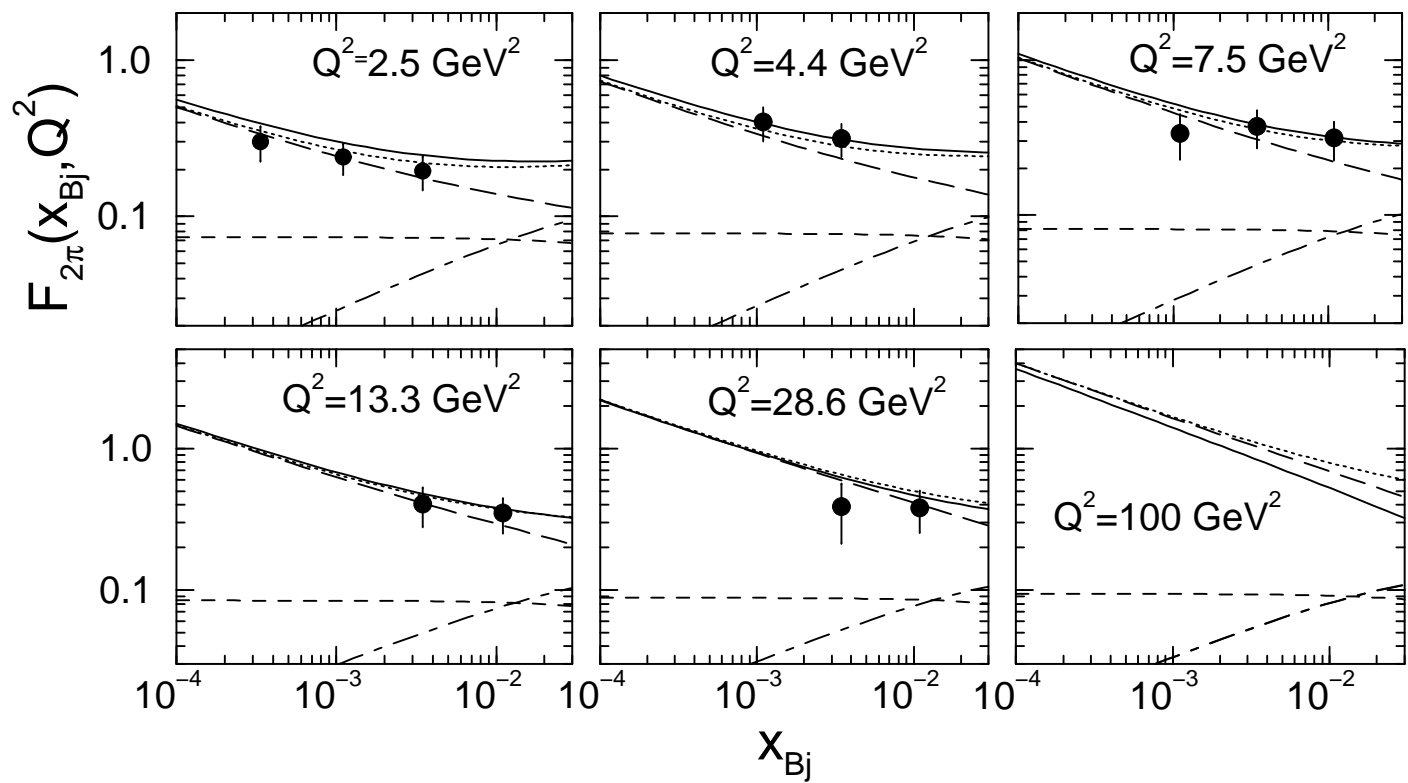
\item[{\bf Fig.1}]
Predictions  from the CD BFKL-Regge factorization for the pion
 structure function
$F_{2\pi}(x_{Bj},Q^2)$ (solid lines).
The experimental data from the H1  Collaboration \cite{F2PI}  are shown by
 full circles.
The different components of $F_{2\pi}(x_{Bj},Q^2)$ are shown:
 the valence contribution (dashed-dotted),
 the non-perturbative soft contribution (dashed), 
the Leading Hard+Soft+Valence Approximation
 $F^{\rm LHSVA}_{2\pi}(x_{Bj},Q^2)$ (dotted) and the sum of the soft and
 leading hard pole contributions without valence component (long dashed).
  
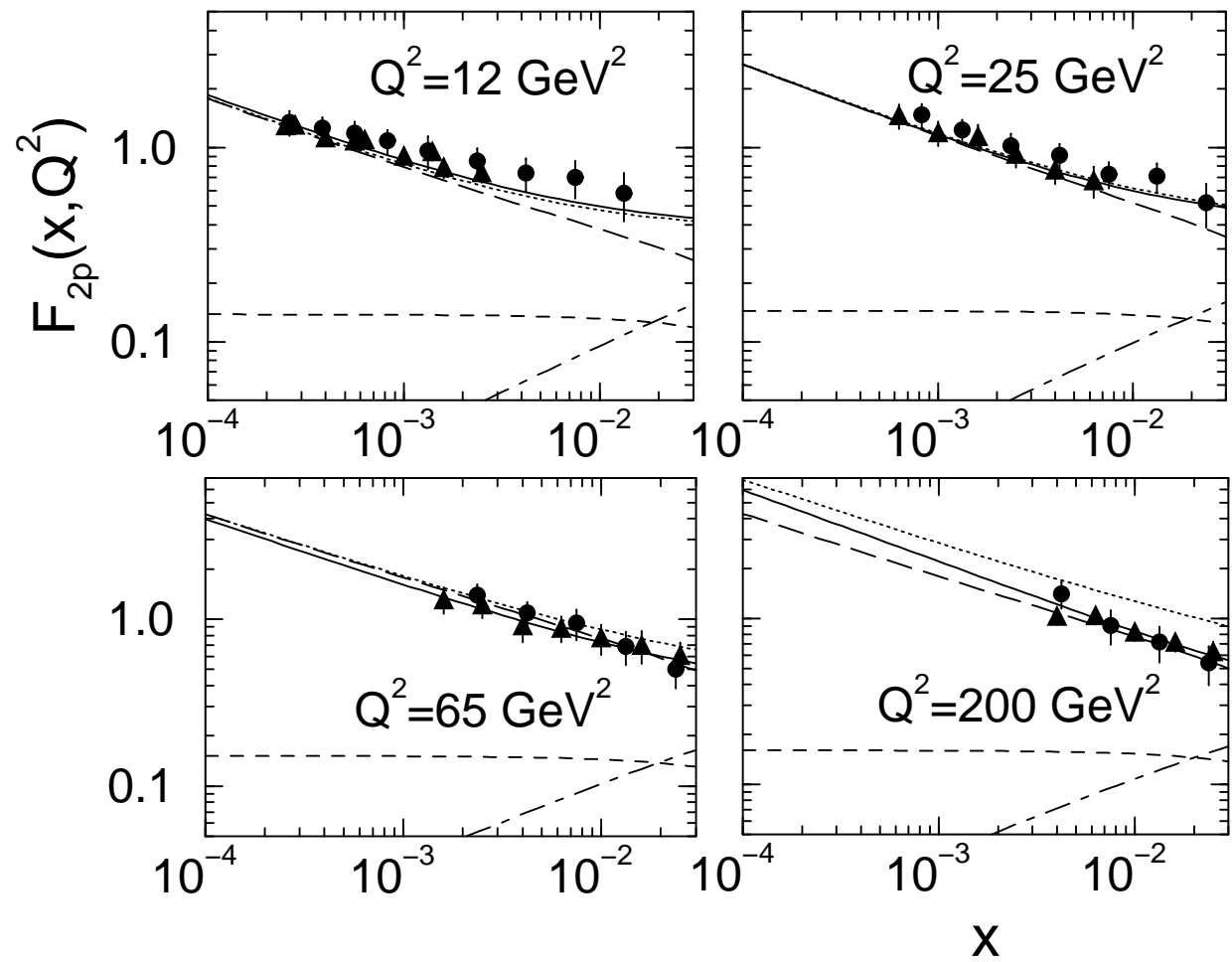
\item[{\bf Fig.2}]
The decomposition into different components of earlier
predictions \cite{JETPLett,DER} from the CD BFKL-Regge factorization for the proton
 structure function
$F_{2p}(x,Q^2)$ (solid lines).
The experimental data from the H1  and ZEUS
  Collaborations \cite{H1F2P,ZEUSF2P}  are shown by circles and triangles, respectively.
The different components of $F_{2p}(x,Q^2)$ are shown:
 the valence contribution (dashed-dotted),
 the non-perturbative soft contribution (dashed),
the  Leading Hard+Soft+Valence Approximation (LHSVA) 
$F^{\rm LHSVA}_{2p}(x,Q^2)$ (dotted) and the sum of the soft and
 leading hard pole contributions without valence component (long dashed).

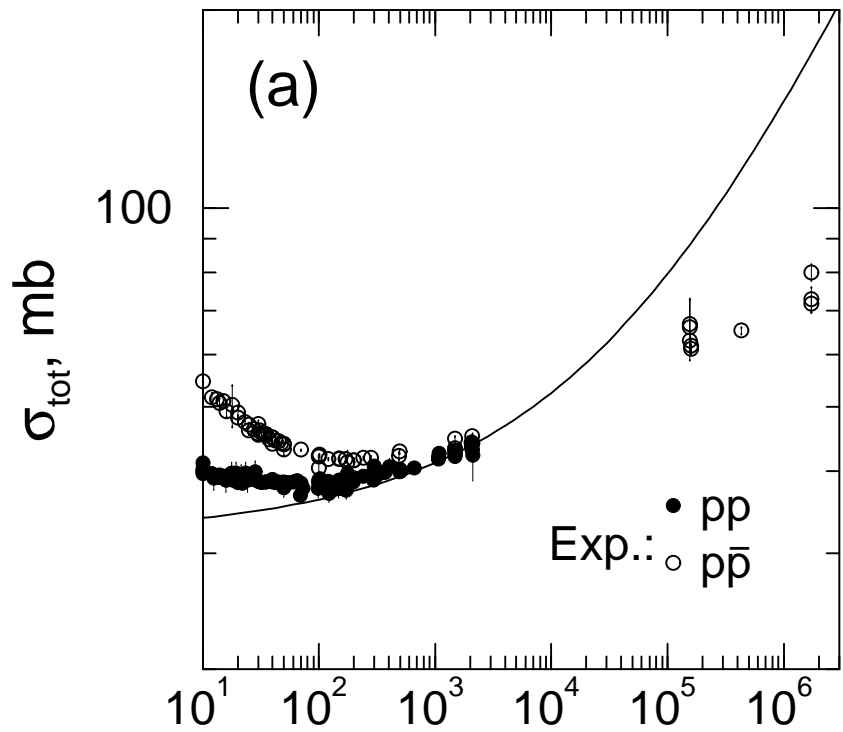
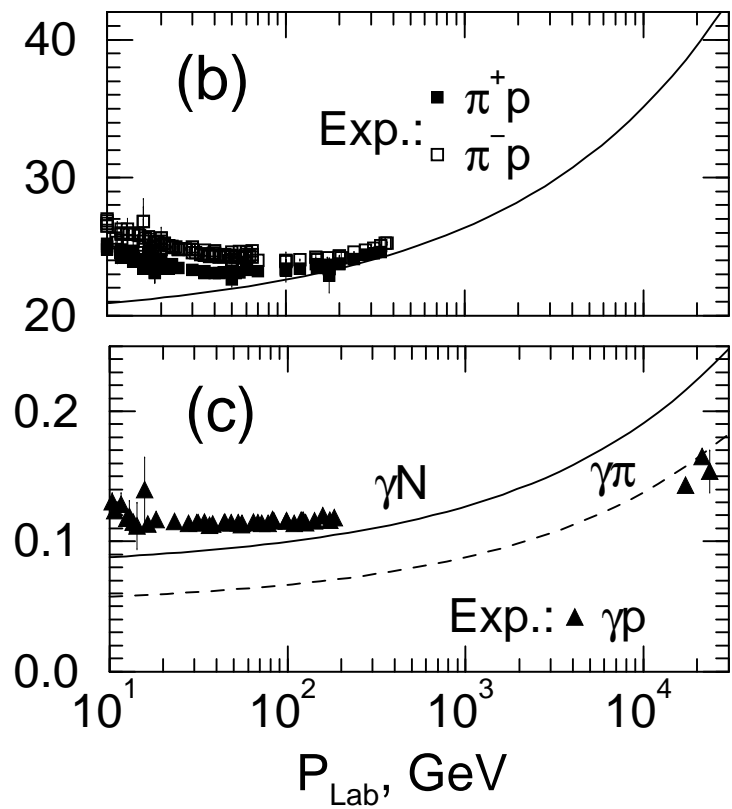
\item[{\bf Fig.3}]
The CD BFKL-Regge factorization evaluation of the vacuum exchange 
 contribution to 
 $\sigma_{\rm tot}^{NN}$ (solid line in box (a)),
 $\sigma_{\rm tot}^{\pi N}$ (solid line in box (b)),
 $\sigma_{\rm tot}^{\gamma N}$ (solid) and
 $\sigma_{\rm tot}^{\gamma\pi}$ (dashed line in box (c)).
The  data points are from \cite{PDG}.

\end{enumerate}

\end{document}